# Visualizing the ultra-structure of microorganisms using table-top extreme ultraviolet imaging

Chang Liu[1,2,3,*,†], Wilhelm Eschen[1,2,3,*], Lars Loetgering[1,2,3,4], Daniel S. Penagos Molina[1,2,3], Robert Klas[1,2,3], Alexander Iliou[5], Michael Steinert[1], Sebastian Herkersdorf[6], Alexander Kirsche[1,2,3], Thomas Pertsch[1,7], Falk Hillmann[5,8], Jens Limpert[1,2,3,7] and Jan Rothhardt[1,2,3,7]

*These authors contributed equally to this work.

[†] Corresponding author, E-mail: liu.chang@uni-jena.de

1. Institute of Applied Physics, Abbe Center of Photonics, Friedrich-Schiller-University Jena, Albert-Einstein-Straße 15, 07745 Jena, Germany

2. Helmholtz-Institute Jena, Fröbelstieg 3, 07743 Jena, Germany

3. GSI Helmholtzzentrum für Schwerionenforschung, Planckstraße 1, 64291 Darmstadt, Germany

4. Present address: Carl Zeiss AG, Carl-Zeiss-Promenade 10, 07745 Jena, Germany

5. Leibniz Institute for Natural Product Research and Infection Biology, Hans Knöll Institute (Leibniz-HKI), Adolf-Reichwein-Str. 23, 07745 Jena, Germany

6. Department of Pharmaceutical Microbiology at the Hans Knöll Institute (HKI), Friedrich-Schiller-University Jena, Winzerlaer Straße 2, 07745 Jena, Germany

7. Fraunhofer Institute for Applied Optics and Precision Engineering, Albert-Einstein-Str. 7, 07745 Jena, Germany

8. Present address: Biochemistry/Biotechnology, Faculty of Engineering, Hochschule Wismar University of Applied Sciences Technology, Business and Design, Philipp-Müller-Str. 14, 23966 Wismar, Germany

**Abstract:**

Table-top extreme ultraviolet (EUV) microscopy offers unique opportunities for label-free investigation of biological samples. Here, we demonstrate ptychographic EUV imaging of two dried, unstained model specimens: germlings of a fungus (*Aspergillus nidulans*), and bacteria (*Escherichia coli*) cells at 13.5 nm wavelength. We find that the EUV spectral region, which to date has not received much attention for biological imaging, offers sufficient penetration depths

for the identification of intracellular features. By implementing a position-correlated ptychography approach, we demonstrate a millimeter-squared field of view enabled by infrared illumination combined with sub-60 nm spatial resolution achieved with EUV illumination on selected regions of interest. The strong element contrast at 13.5 nm wavelength enables the identification of the nanoscale material composition inside the specimens. Our work will advance and facilitate EUV imaging applications and enable further possibilities in life science.

**Introduction:**

Visual information obtained through microscopy is vital for our understanding of the microbial world. High-resolution imaging of microorganisms has important implications for pharmaceutics, medicine, and biological research. Numerous biological imaging techniques have been explored extensively and may be classified by the energy spectrum of the radiation involved. Three possible categories are 1) optical (e.g., confocal, multiphoton, fluorescence microscopy), 2) X-ray, and 3) electron microscopy (EM). Super-resolution visible fluorescence microscopy[1-3] enables specific chemical contrast for chosen fluorescent labels in cellular and molecular biology. Cryogenic-EM provides near-atomic resolution for macromolecular structure determination. Due to the relatively small penetration depth of the electron[4-7], EM is complemented by X-ray microscopy, allowing for microscopy and tomography at mesoscopic length scales[8].

With the advent of third-generation synchrotron sources, X-ray microscopy has matured, offering large penetration depths and nanometer-scale three-dimensional resolution. Hard X-rays with multi-keV photon energies can be used to stimulate X-ray fluorescence from most biologically relevant trace elements in cells and tissues[9-11]. However, hard X-rays typically exhibit poor absorption contrast for whole-cell structural imaging.

A much higher natural contrast between water and biological macromolecules like proteins is found in the soft X-ray region, in the so-called water window at energies between the carbon and oxygen absorption edges at 290 eV and 540 eV. Here, microscopes have been implemented to image the whole, dehydrated[12,13], or frozen-hydrated cells[14-18], allowing for the visualization of cellular and subcellular features in their native state.

Meanwhile, recent progress in coherent X-ray and extreme ultraviolet (EUV) sources has led to a growing impact of so-called lensless imaging techniques, also known as coherent diffraction imaging (CDI). Unlike lens-based imaging techniques, CDI avoids absorptive losses and exceeds the resolution limit inherent to image-forming optics[19]. Ptychography[20], a scanning version of CDI, records a sequence of diffraction patterns from overlapping illumination regions on extended objects. This approach computationally retrieves amplitude and phase information of the object and illumination wavefront, which allows the extraction of quantitative information about the material composition.

Due to the high brightness and coherence requirements, current lensless X-ray microscopes rely on synchrotron-radiation and free-electron lasers[21-26], thus limiting the widespread access to CDI techniques. However, in the EUV spectral region, coherent radiation with steadily increasing photon flux and brilliance can be obtained utilizing high harmonic generation (HHG)[27] driven by femtosecond high-average power lasers. Within the past decade, these sources have seen tremendous progress in terms of photon flux and stability[28], which enables nanoscale coherent imaging on a table-top[29] with applications ranging from reflectometry[30,31] and wavefront sensing[32,33] to material sciences[34,35]. EUV radiation provides very high absorption- and phase contrast, moderate penetration depth, and still possesses short enough wavelengths to resolve sub-20 nm structures[35,36]. Further, a variety of atomic resonances across the periodic table lies in this spectral region, enabling chemically-resolved imaging[37], further complementing the capabilities of X-rays and electrons.

For biological applications the EUV spectral region has remained a widely unexplored territory. A first demonstration at 29 nm wavelength imaged the outer contours of mouse neurons with sub 80 nm lateral resolution[38], while the intracellular structures remained unexplored. This was due to the limited penetration depth (sub-100 nm) into typical biological materials at this particular wavelength.

In this work, we present HHG-based EUV ptychographic imaging of microorganisms at a significantly shorter wavelength of 13.5 nm (92 eV photon energy). Here, dried and unstained germinating conidia of the filamentous fungus *Aspergillus nidulans* (*A. nidulans*) and the cells of the bacterium *Escherichia coli* (*E. coli*) are investigated. A large field-of-view (FOV) overview image of the whole sample is provided by ptychography with infrared illumination, which facilitates the identification and pre-selection of relevant regions of interest (ROIs). Subsequently, EUV ptychography provides complex transmission images of the specimens with a half-period spatial resolution of 58 nm. Information about the interior material composition of the investigated microorganisms is extracted from these EUV images based on the analysis of the scattering quotient[39]. This scattering quotient micrograph uncovers the composition of the sample averaged along the propagation direction in each pixel of the image. In both investigated samples, different biological compositions have been obtained and successfully assigned to the internal functional units of the respective microorganisms.

**Results**

*Experimental setup*

The recent progress in ultrashort fiber laser systems enables average powers in the kilowatt range while offering millijoules of pulse energy[40]. Driving the HHG process with such lasers results in coherent EUV radiation, nowadays exceeding powers of 10 mW at 47 nm wavelength[41]. For shorter wavelengths, the conversion efficiency and thus the available EUV power drops off rapidly. As a result, the shortest wavelength at which HHG sources can provide sufficient average power for high-resolution imaging is currently found in the region surrounding 13.5 nm wavelength[35,36].

Here, we use a designated home-built high-power, few-cycle fiber laser system to generate a broadband EUV continuum with a state-of-the-art photon flux of $7 \cdot 10^9$ ph/s/eV at 13.5 nm wavelength[42]. A detailed description of this EUV generation source can be found in the *Materials and Methods* section. The basic setup of the ptychographic microscope has been described elsewhere[35]. This setup has been extended to enable dual-wavelength imaging in the infrared as well as in the EUV. A schematic overview is shown in Figure 1**a** and **b**. The reflective optics in the beam path provide sufficient reflectivity not only at their design wavelength in the EUV (13.5 nm) but also at the infrared wavelength of the fundamental laser beam (1030 nm). We can thus utilize either of the two wavelengths for the illumination of the sample.

Structured illumination is used for both the IR and the EUV illumination to enrich the spatial-frequency spectrum in the illumination, thus enhancing the signal-to-noise ratio (SNR) and spatial resolution of ptychographic reconstructions[43]. For this purpose, two amplitude masks with diameters of 300 µm for IR illumination and 10 µm for EUV illumination were fabricated with a separation of 705 µm in a single long-slit $Si_3N_4$ membrane, which is mounted on a 2D positioner (Figure 1**c**) allowing to switch between the two masks. Both amplitude masks have a spiral shape to tailor the structure of the IR and EUV illumination. The microorganisms to be imaged, are prepared on a $Si_3N_4$ membrane that is located a few hundred microns downstream of the illumination masks. They are cultured and plated on the $Si_3N_4$ membrane following standard protocols avoiding stains or chemicals (see *Materials and Methods* section). This allows for a simple and straightforward preparation but results in sample dehydration which can alter the native morphology (e.g., reduction of cell size) and chemical distribution[44].

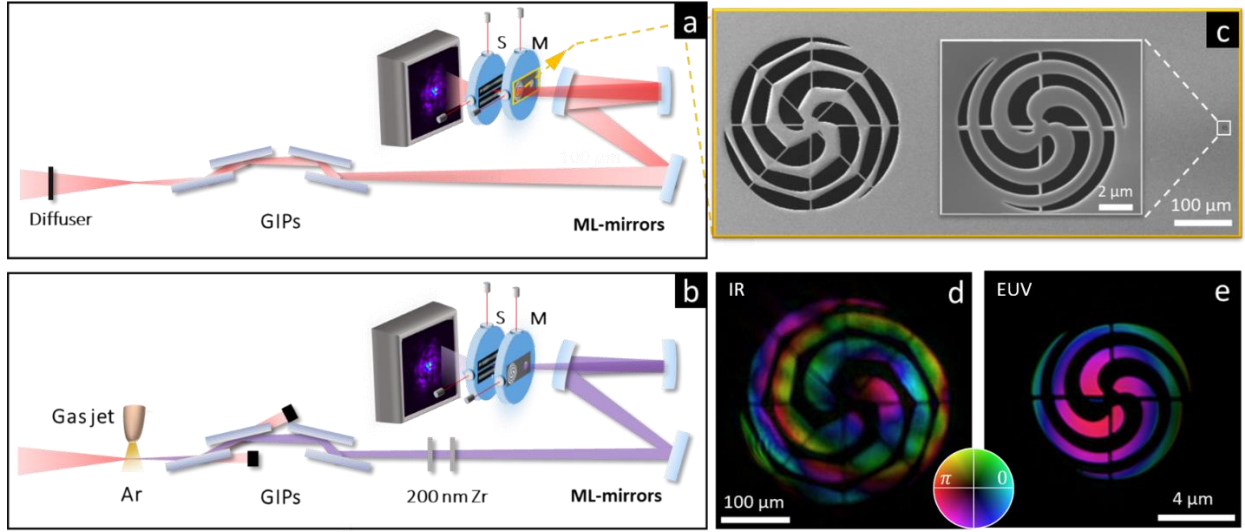

**Figure 1: Schematic setup of the position-correlated EUV/IR ptychographic microscope. a** and **b** show the IR and EUV ptychographic setup. GIPs: grazing incident plates. ML mirrors: multilayer mirrors. M: mask. S: sample. **c** Scanning electron microscope (SEM) image of the amplitude-only masks which are used to structure the IR and EUV beam respectively. For the IR mask (left spiral in **c**) a diameter of 300 µm was used and for the EUV mask (right in **c**) a diameter of 10 µm was used. The inset in **c** shows an enlarged SEM image of the EUV mask for illustration. The distance between the two masks is 700 µm. **d** Reconstructed complex probe back-propagated to the mask plane in IR scan. **e** Reconstructed probe back-propagated to the mask plane in EUV scan.

*Position-correlated IR- and EUV- ptychography*

EUV ptychography can provide nanoscale resolution, but also suffers from a small FOV, which scales proportionally to the deployed wavelength. By using the more powerful driving IR laser, an overview image of the sample with a millimeter-scale FOV and micrometer-scale resolution can be captured. Here, we combined the advantages of both spectral regions and demonstrate position-correlated IR- and EUV-ptychography.

The method is verified by imaging clusters of dried germlings of *A. nidulans* conidia, prepared on a $Si_3N_4$ membrane. For recording the IR images, a small fraction of the IR beam (~ 0.5 mW) centered at 1030 nm is selected by ND filters. A diffuser was placed in the beam path to get a more diverse beam, while the HHG gas jet and two 200 nm Zr filters are removed (see Figure 1 **a** and **b**). The whole membrane area was scanned by an elongated spiral grid, resulting in a FOV as large as $0.9\ mm^2$.

The complex transmission image of the probe is back-propagated numerically to the mask plane (Figure 1**d**) and matches the SEM image of the mask (Figure 1**c**). The reconstruction (see *Materials and Methods* section) of the complex transmission of the sample membrane at IR wavelength is shown in Figure 2**a**. Despite the low spatial resolution (1 ~ 2 µm in the selected region, see supplement Figure S2), clusters of *A. nidulans* are visible, which aggregate in some areas and tend to be sparse in others.

In the next step, we performed a high-resolution investigation of the regions of interest (ROIs) with clusters of *A. nidulans* using the generated EUV illumination. The beam is spectrally selected by the reflective optics at around 13.5 nm wavelength and spatially filtered by the mask with a diameter of 10 µm (see inset of Figure 1**c**). The complex probe in the mask plane was again obtained via numerical backpropagation and matched well the structure of the mask (Figure 1**e**).

Since the IR and EUV beams are collinear, the position and inclination of the sample relative to the illumination remain constant, ideally the two scans share the same coordinates. As a result, the positions of both scans are intrinsically correlated, and thus pre-selected ROIs can be directly obtained from the IR image for navigating EUV measurements. We found that the position shear between these images is around 1 µm, which is comparable to the pixel size of the IR image. Details on the analysis of the position correlation are provided in the supplement.

Fungal clusters containing several *A. nidulans* hyphae were investigated using various scan grids with the size of 40 ~ 60 µm. The reconstructed micrographs are shown in Figure 2**b**, **c**, and **e**. In these images, hue and brightness encode phase shift and transmissivity, respectively. For further investigation, regions containing a single hypha were selected at smaller raster sizes and imaged.

The reconstructions are shown in Figure 2**d** and **f**. The internal structures are visible, including tubular structures inside the hypha, as shown in Figure 2**d**.

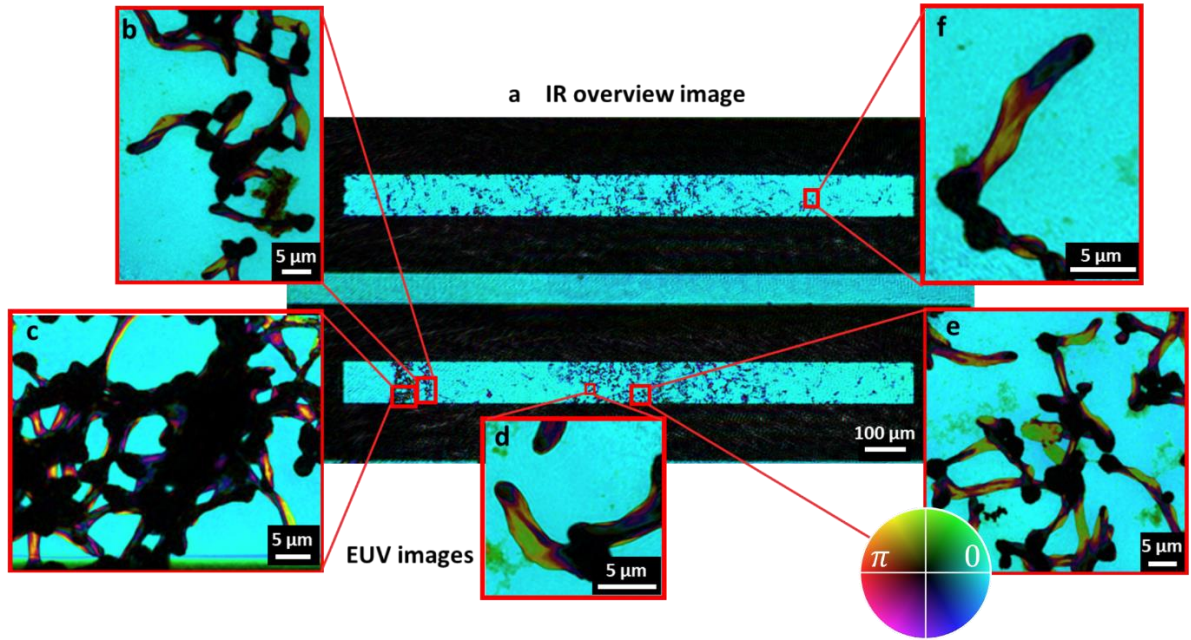

**Figure 2 IR and EUV complex reconstruction of *A. nidulans*. a** Reconstruction of the entire $Si_3N_4$ membrane using the fundamental IR beam. The FOV of the reconstruction is 0.9 mm². **b - f** show the reconstructed complex EUV images measured using a narrow spectrum from the generated high-order harmonics of the driving IR beam. In **b**, **c**, and **e**, clusters of hyphae are observed, while **d** and **f** show the structure of a single hypha. The image brightness represents the transmitted amplitude, and the hue represents the phase.

To create a high-resolution image, a second, high-dynamic-range (HDR) measurement of an area that contains a single hypha (as displayed in Figure 2**f**) was carried out. The corresponding half-pitch resolution was evaluated by a Fourier ring correlation (FRC) to be 58 nm (see supplement Figure S2). The dose in this HDR measurement was estimated[8] to be $7.4 \times 10^4$ Gy, which is slightly lower than the dose estimated in previous synchrotron work to image dry specimens in the water window[26]. This value is far below the threshold of radiation damage effects in biological samples[8] (see supplement Figure S4) and no structural changes were observed during the measurements. In addition, we estimated the theoretical dose required to image a feature (representative biological molecules, i.e., carbohydrate, protein, and others) with the voxel size of 100 nm in a dehydrated cellular environment based on the model reported in prior publications[45,46].

The detailed dose calculations, found in the supplement, corroborate our resolution estimated (58 nm half-period resolution) from the contrast analysis point of view.

*Chemically-sensitive EUV imaging of a single A. nidulans hypha*

For further analysis, the reconstructed amplitude and unwrapped phase of the high-dynamic-range image in Figure 2**f** are separately shown in Figure 3**a** and **b**, respectively. In this close-up image, the structure of a single hypha is depicted in more detail. At the lower end of the germling, a round shape is apparent which does not transmit EUV radiation and can be attributed to the conidiospore from which the germling has grown. From this, a tubulus with a diameter of approximately 2 μm is growing with a tip at its end. It appears that the phase (Figure 3**b**) at the tip reaches a higher value, which indicates either a change in material or more overall penetrated material. In the next step the scattering contrast, which is related to the cross-section for coherent scattering[47] is calculated from the reconstructed amplitude and unwrapped phase and is shown in Figure 3**c**. This quantity combines amplitude and phase contrast in a single real number and therefore gives a measure for the total scattering integrated over the thickness of the sample. Analysis of the scattering contrast indicates that the hypha can be divided into three areas, which are labeled here by **i**, **ii**, and **iii**. It appears that in region **i** the scattering contrast is the lowest while in region **ii** the scattering contrast rises and experiences a peak in region **iii**. We believe that at the top of the hypha (region **iii**) the so-called 'Spitzenkörper' (SK) is visible. The SK is rich in phospholipid vesicles and proteinaceous filaments. It represents the cellular growth point of filamentous fungi[48].

A detailed analysis of the material composition can be obtained by the projected scattering quotient $f_q$ (Figure 3**d**), which is calculated from the reconstructed amplitude and phase[23]. The scattering quotient is independent of the thickness and density of the contained materials and depends solely on the chemical composition. In the 2D images presented here, it represents the average elemental composition of the various cellular components along the projection direction in each image pixel.

The reconstructed scattering quotient is subsequently compared to the theoretical values[49], allowing for the identification of the contained materials (see *Materials and Methods* section). The scattering quotient values for the important components of biological samples (carbohydrate, phospholipids, lipids, nuclear acid, and protein) are given in Table 1. Since these values are well separated, the dominating components of different subcellular units can be recognized by scattering quotient analysis.

Histograms of scattering quotients for the regions 1, 2, and 3 marked in Figure 3**d** are shown in Figure 3**f**. The well-separated modes histograms indicate that the regions have different chemical compositions. In region 1 the average scattering quotient is calculated to be 3.3, which is close to the value of carbohydrates ($f_q = 3.5$). Since the hypha was in a dried state, it is assumed that this region consists mainly of the cell wall. From the literature, it is well known that the cell wall of *A. nidulans* contains mainly glucans and thus carbohydrates[50]. In region 2 the average scattering quotient increases to a value of $f_q = 4.0$. Within this region a significant amount of proteins and lipids are present, indicating the presence of subcellular components (e.g., microtubules and vesicles). Most interestingly, at the tip of the hypha (region 3) the scattering quotient reaches a peak value of $f_q = 4.5$. The high scattering quotients indicate the presence of the SK, which is well known for its high phospholipid concentration[51]. The SK is surrounded by a low scattering quotient value ring (Figure 3**e**), which may be attributed to macrovesicles containing glucans oligomers required for cell wall biosynthesis[52].

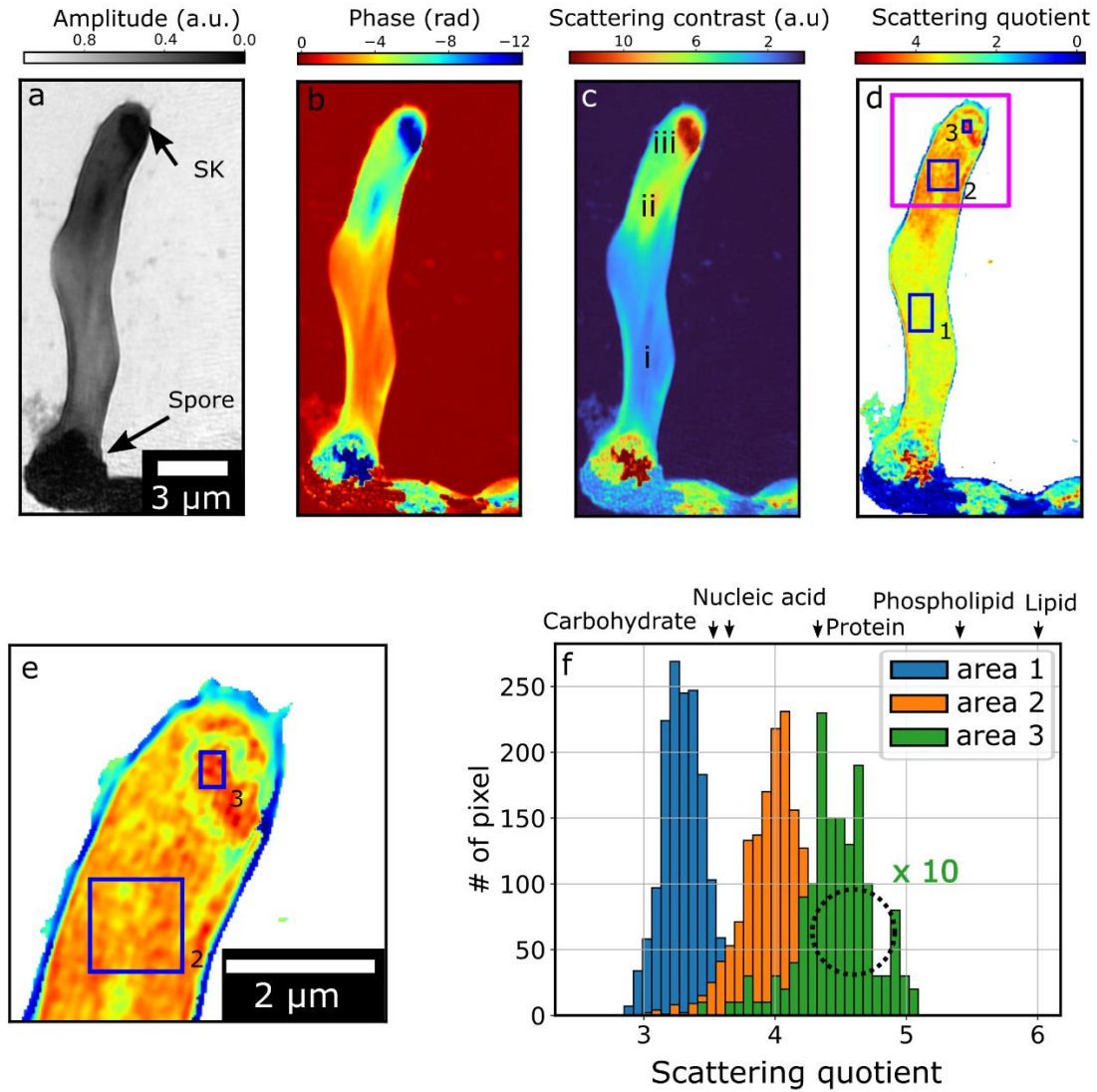

**Figure 3 High-resolution EUV image of *A. nidulans*. a** and **b** show the reconstructed amplitude and unwrapped phase respectively. At the bottom, the spore is visible, which shows low transmission. While at the center the tubulus is visible and at the top, the high phase shift at the top indicates the Spitzenköper (SK). In **c** the scattering contrast is shown, where three regions are indicated by **i**, **ii**, and **iii**. **d** shows the scattering quotient. **e** shows the magnified region indicated in **d** by the pink box. **f** Histogram of the scattering quotient in the blue boxes in **d** and **e** by the blue boxes labeled 1, 2, 3. The number of pixels for area 3 is multiplied by a factor of 10 to facilitate comparison.

|              | H  | C  | N  | O  | P | S | $f_q$ (13.5 nm) |
|--------------|----|----|----|----|---|---|-----------------|
| **Protein**      | 50 | 30 | 9  | 10 | 0 | 1 | 4.3             |
| **Nucleic acid** | 51 | 39 | 15 | 25 | 4 | 0 | 3.6             |
| **Lipid**        | 98 | 55 | 0  | 6  | 0 | 0 | 6.0             |
| **Phospholipid** | 79 | 42 | 1  | 8  | 1 | 0 | 5.4             |
| **Carbohydrate** | 10 | 6  | 0  | 5  | 0 | 0 | 3.5             |

**Table 1** Chemical compositions and their corresponding scattering quotient ($f_q$) for relevant biochemical components at a photon energy of 92 eV (13.5 nm) taken from the literature[49]. The values under elements correspond to the relative proportions of the elements in the chemical formula of typical biological compounds, e.g., $C_{10}H_6O_5$.

*Ptychographic EUV imaging of E. coli cells*

In the next step, *E. coli* cells, widespread as a prokaryotic model organism, were investigated. As compared to *A. nidulans* germlings, *E. coli* cells are significantly smaller with a typical size of 1 μm × 2 μm which reduces the overall scattering signal. Therefore, at each scan position, three diffraction patterns with three different exposure times (0.3s, 3s, 9s) were captured and fused into a single high-dynamic range image. The reconstructed complex specimen micrograph is shown in Figure 4**a**. The corresponding half-period resolution was estimated by an FRC to be 74 nm (see supplement). Next to multiple clustered and single *E. coli*, the reconstruction shows a dense meshwork of the extracellular matrix between single cells in the center. For further analysis, a region that appears to contain two *E. coli* cells was identified (red box in Figure 4**a**). The scattering quotient was calculated for the corresponding region and is shown in Figure 4**b**. Low values of the scattering quotient are apparent at the boundary of the cell which indicates that this area mainly consists of carbohydrates, indicating the location of the cell wall. However, it is notable that there seems to be no cell wall between the two objects, indicating that the two entities correspond to a single cell in the stage of cell division. In the center of both bodies the value of the scattering quotient increases, which is consistent with the presence of proteins, DNA (nucleic acid), and lipids.

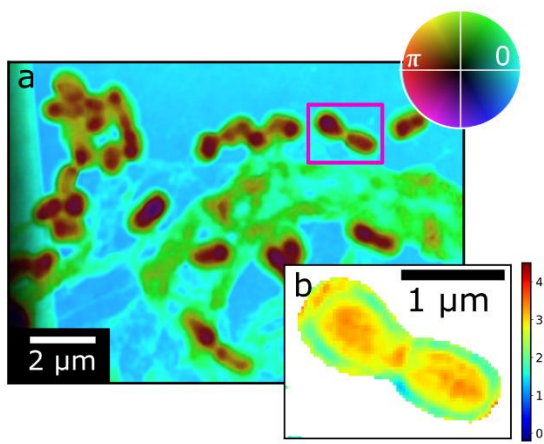

**Figure 4 High-resolution EUV ptychography reconstruction of *E. coli* bacteria a** shows the complex transmission of multiple *E. coli* bacteria prepared on a $Si_3N_4$ membrane, where the image brightness represents the transmitted amplitude, and the hue represents the phase. For a single *E. coli* cell in the state of cell division, which is indicated by a pink box in **a**. the scattering quotient is calculated and shown in **b**. Low values of the scattering quotient correspond to the cell wall, which is dominated by carbon hydrates, while the large scattering values correspond to proteins, lipids, and nucleic acid.

## *Discussion*

We demonstrated high-resolution chemically-sensitive EUV imaging of microorganisms using a table-top high harmonic source. Compared to previous work[38], the 92 eV EUV photons provide a significantly longer penetration depth, enabling imaging of intracellular features of comparably thicker samples in the µm-range. IR- and EUV ptychography provides a combination of millimeter-squared FOV overview and high-resolution imaging of smaller regions of interest. Sub-60 nm spatial resolution is achieved and enables uncovering of the structural information of subcellular features. Moreover, an analysis of the quantitative amplitude and phase information obtained by EUV ptychography allows drawing conclusions on the chemical composition of the biological samples. In this way, subcellular components such as the SK of *A. nidulans* and the cell wall have been identified.

Our work shows that EUV ptychographic imaging provides unique capabilities and can complement existing biological imaging modalities. First, it offers a high amplitude and phase contrast compared with X-ray microscopy for dried cells (shown in supplement Figure S4**b**) with straightforward sample preparation. Second, due to the high refractive index contrast in the EUV spectral region, the required dose absorbed by the biological samples to achieve similar resolution is lower as compared to X-ray (i.e., in water window) and electron-based microscopy[45]. We estimate that, with an optimized microscope and a higher photon flux source, achieving sub-20 nm half-pitch resolution will be possible with a radiation dose of $< 3 \times 10^5$ Gy, which is below the radiation damage threshold (shown in supplement Figure S4**d**). Third, ptychography enables to retrieve excellent scattering contrast and averaged elemental distributions can be analyzed without the need for labeling or staining.

The imaging performance mentioned above comes in a table-top format, which will allow for widespread use in life science and potentially even in clinical environments. We believe that our work demonstrates the strengths of EUV ptychography to perform high-resolution, chemical-sensitive biological imaging.

In the future, a combination with other microscopy techniques such as nonlinear-, fluorescence- or coherent anti-stokes Raman scattering microscopy with visible and infrared illumination appears feasible with only minor modifications of the presented setup. This intrinsically promises position-correlated multi-modal imaging of functional, cellular, and subcellular structures in a single device, even experiments that have not been proposed yet can be carried on.

**Materials and Methods**

*High harmonic generation*

The high harmonic generation process is driven by a fiber-based chirped-pulse amplifier operating at a central wavelength of 1030 nm. The amplified pulses are compressed to < 7 fs by cascaded noble-gas-filled hollow-core fibers with a residual average power of 30 W at a repetition rate of 75 kHz and pulse energy of 400 µJ. These few-cycle pulses are directed into a vacuum chamber and focused into a 700 µm diameter gas jet with a backing pressure of 600 mbar argon. A broad EUV continuum is generated, with a photon flux of $7 \times 10^9$ photons/s/eV at 92 eV. The separation of the generated harmonics from the high-power driving laser is realized by four grazing incidence plates (GIPs). Afterward, the residual IR laser is fully blocked utilizing two 200 nm zirconium (Zr) foils, which allow a transmission window between 70 eV and 120 eV. More details on the HHG source can be found in our earlier work[42].

*Data collection and processing*

In the experimental setup, both the sample and mask were mounted onto individual 2D positioning systems with active stabilization by means of a laser interferometer-based feedback loop[35]. For this purpose, the mask and sample were placed on XYZ piezo positioners (SLC 1740, SmarAct GmbH) and the positions were measured and actively stabilized by a laser interferometer (Picoscale, SmarAct GmbH). This setup improved the long-term stability of the positioners, which is crucial for ptychography. The diffraction patterns were recorded by a CCD detector (Andor iKon-L, 13.5 µm pixel size, 2048 × 2048) at a sample-detector distance of 31 mm, leading to a detection numerical aperture (NA) of 0.41 when the full chip of the detector area is used. To minimize the thermal noise throughout data acquisition, the camera was cooled down to -60 °C.

All the microbial samples were scanned using the Fermat spiral scan grids[53]. For the IR scan on the *A. nidulans* sample (Figure 2**a**), the spiral grid was adjusted to fit the elongated geometry of the entire membrane. The scan step was set to 40 µm. The CCD pixels were read out at a rate of 1 MHz and the on-chip binning was set to 4 × 4 to reduce the readout time. The IR scan with 850 scan positions and a FOV of about 0.9 mm² took 62 minutes, mainly due to the slow readout of the employed CCD detector. Note that in the future recording such overview images will be feasible within a few seconds by combining commercially available rapid readout EUV sCMOS detectors[56] with the fly-scan method[57].

For the EUV scans on the individual ROIs (Figure 2**b ~ f**) of *A. nidulans* and *E. coli*, the scan step between adjacent positions was chosen between 1 µm and 2 µm, according to the requirements of the FOV. The on-chip binning was set to 2 × 2. Here for *A. nidulans*, two diffraction patterns with a short (0.3 s) and a long exposure time (4 s) were acquired for each scan position and combined into a single high dynamic range diffraction pattern. For *E. coli*, 119 EUV diffraction patterns were recorded with exposure times 0.3 s, 3 s, and 9 s, and fused into a single HDR pattern. A table with the measurement parameters of each image is shown in the supplement.

First, the raw diffraction patterns were background subtracted. After these preprocessing procedures, the object and probe were reconstructed by standard ptychography reconstruction algorithms, which were performed using the GPU-accelerated package of ptylab[54]. An axial position calibration algorithm (zPIE[55]) was used to estimate the sample-detector distance. Subsequently, an accelerated gradient solver (mPIE[58]) was used to reconstruct the object and probe combined with the mixed-stated forward model[59] to account for mode de-coherence effects, such as high-frequency sample vibrations[60], background, detector point-spread, and a finite spectral bandwidth[61] in the illumination. To compensate for slow probe wavefront variations on a time scale longer than the exposure time, the orthogonal probe relaxation (OPR) method[62] was applied to all illumination modes, which is called *mixed-state orthogonal probe relaxation* (m-s OPR)[35]. For the quantitative analysis of *A. nidulans* and *E. coli* samples, four mixed states each consisting of 4 OPR modes were modeled, resulting in a total of 16 probe modes. The resulting reconstructions contain fewer artifacts and become more quantitatively reliable for the material composition analysis as compared to simpler forward models.

*Scattering quotient and scattering contrast analysis*

The atomic scattering factor $f = f_1 + if_2$ is tabulated for all elements and can be found in the literature for the EUV and X-ray spectral range up to 30 keV[49]. To calculate the atomic scattering factor $\bar{f} = \bar{f_1} + i\bar{f_2}$ for a compound (e.g., carbohydrate $C_6H_{10}O_5$) the mean atomic scattering factor for the real and imaginary parts has to be calculated, which accounts for the stoichiometric weight $s_i$ of all constituent elements[8].

$$\bar{f_1} = \sum_i s_i f_{1,i}$$

$$\bar{f_2} = \sum_i s_i f_{2,i}$$

The theoretical scattering quotient is subsequently given by the ratio of the resulting real and imaginary parts. Experimentally, the complex scattering quotient can be calculated from the reconstructed complex transmission function via the ratio of the unwrapped phase $\phi(x,y)$ and the natural logarithm of the reconstructed amplitude $|O(x,y)|$[23].

$$\bar{f_q} = \frac{\bar{f_1}}{\bar{f_2}} = \frac{\phi(x,y)}{\ln(|O(x,y)|)}$$

Since ptychographic reconstructions are invariant to a scaling of the amplitude and global phase shift, the reconstructed complex transmission of the object is referenced to a known vacuum area, where an amplitude of unity and zero phase shift is assumed.

Similar to the scattering quotient, the scattering contrast[47] can be defined by

$$c = \frac{\sqrt{\delta^2 + \beta^2}}{\lambda^2}$$

where $\delta$ and $\beta$ correspond to the deviation from the unity of the real and imaginary part of the refractive index $n = 1 - \delta - i\beta$. Starting from the reconstructed amplitude $|O(x,y)|$ and phase $\phi(x,y)$, the complex reconstruction can be converted to a real-valued image that combines phase and amplitude contrast

$$\psi(x,y) = \sqrt{\ln(|O(x,y)|^2 + \phi(x,y)^2} = 2\pi * c * t(x,y)$$

where $t(x,y)$ corresponds to the 2-dimensional thickness map of the sample. Comparing with the cross-section for elastic scattering[8] ($r_e$ corresponds to the classical electron radius)

$$\sigma_{el} = \frac{8}{3} * \pi r_e^2 (f_1^2 + f_2^2)$$

it is evident that the calculated scattering contrast value c directly relates to the elastic scattering cross-section. Therefore, $\psi(x,y)$ gives a measure for the total elastic scattering (i.e. diffraction) in each pixel integrated along the projection direction.

*Sample preparation and characterization*

Both, *A. nidulans* germlings and *E. coli* were cultivated in 8-well plates (Sarstedt) and adhered to 50 nm thick $Si_3N_4$ membranes. For this purpose, $10^3$, $10^4$, and $10^5$ conidia of *A. nidulans* FGSC4 were inoculated in single wells with 3-5 ml of Czapek-Dox broth (BD Difco™) and incubated at 30 °C for 16 hours. The germlings were then carefully washed twice with $H_2O$ to eliminate remnants of any growth media and non-adherent germlings.

*E. coli* K12 was inoculated from overnight cultures and grown in LB-medium overnight at 30° C. It was then carefully washed with distilled $H_2O$ to remove medium and non-adherent bacteria. The residual $H_2O$ was aspirated, and the air-dried samples were imaged.

*Mask preparation*

The probe mask is used to create structured illumination for ptychography. A 50 nm thick $Si_3N_4$ membrane with an effective area of $1500 \times 500\ \mu m^2$ was used as a ground for mask fabrication.

After coating it with 50 nm of copper (thermal evaporation) from the backside to support charge dissipation during the structuring process, a focused Ga+ ion beam (FEI Helios G3 UC, 30 keV, 21 nA for IR mask, 230 pA for EUV mask) was scanned over the Si3N4 surface to etch the desired aperture through the membrane and the copper layer. For this purpose, a black and white bitmap was used to define the structure consisting of 1024x1024 pixels with a defined pitch by toggling the exposure time between 0 and the given dwell time for black and white pixels, respectively. The pitch was 300 nm for the IR mask and 12 nm for the EUV mask, and the dwell time was 5 ms for the IR mask and 200 µs for the EUV mask, respectively. The writing was done within a single pass. Afterwards, an additional 150 nm of copper was deposited on the backside to achieve a final absorber thickness of 200 nm Cu + 50 nm Si3N4. Finally, the aperture shape was confirmed using scanning electron microscopy (SEM, see Figure **1**c). Here it can be seen that the intrinsic stresses of the deposited copper layer lead to a deformation of the IR mask from the ideal spiral shape.

**Data Availability**

The data that support the plots within this paper and other findings of this study are available from the corresponding author upon reasonable request.

**Acknowledgments**

The research was sponsored by a Strategy and Innovation Grant from the Free State of Thuringia (41-5507-2016), the Innovation Pool of the Research Field Matter of the Helmholtz Association of German Research Centers (project FISCOV), the Leibniz Research Cluster InfectoOptics (SAS-2015-HKI-LWC) , the Thüringer Ministerium für Bildung, Wissenschaft und Kultur (2018 FGR 0080), the Helmholtz Association (incubator project Ptychography 4.0) and the Fraunhofer-Gesellschaft (Cluster of Excellence Advanced Photon Sources). A. I. and F. H. acknowledge the support by the German Research Foundation (Deutsche Forschungsgemeinschaft, DFG) under Germany's Excellence Strategy – EXC 2051 – Project-ID 390713860. S.H. is supported by the German Research Foundation (Deutsche Forschungs-gemeinschaft, DFG) – SFB 1127/2 ChemBioSys – 239748522. We thank Dirk Hoffmeister for his help with a fruitful contribution to the manuscript.

**Author Contributions**

C.L., W.E., and R.K. performed the imaging experiments. C.L., W.E., L.L., D.S.P.M, and J.R. analyzed the data. C.L. and L.L. designed the probe mask. M.S. fabricated the mask and acquired the electron microscope images. S.H. contributed to the conceptualization of model bio-specimens measurement. A.I. and F.H. prepared the biological specimens: germlings of the fungus (*A. nidulans*) and the bacterium (*E. coli*). All authors discussed and contributed to the interpretation of the results and the writing of the manuscript. J.R., J.L., F.H., and T.P. supervised the project.

**Competing interests**

The authors declare that they have no known competing financial interests or personal relationships that could have influenced the work reported in this paper.

**References**

1. Balzarotti, F. et al. Nanometer resolution imaging and tracking of fluorescent molecules with minimal photon fluxes. *Science* **355**, 606-612 (2017).
2. Betzig, E. et al. Imaging intracellular fluorescent proteins at nanometer resolution. *Science* **313**, 1642-1645 (2006).
3. Hell, S. W. & Wichmann, J. Breaking the diffraction resolution limit by stimulated emission: stimulated emission depletion microscopy. *Opt. lett.* **19**, 780–782 (1994).
4. Koster, A. J. & Kluperman, J. Electron microscopy in cell biology: Integrating Structure and Function *Nat. Rev. Mol. Cell Biol.* **4**, SS6-SS9 (2003).
5. Kourkoutis, L. F. et al. Electron microscopy of biological materials at the nanometer scale. *Annu. Rev. Mater. Res.* **42**, 33-58 (2012).

6. Ross, F. M. *Liquid Cell Electron Microscopy* (Cambridge Univ. Press, 2016).

7. Vénien-Bryan C. et al. Cryo-electron microscopy and X-ray crystallography: complementary approaches to structural biology and drug discovery. *Acta Crystallogr., Sect. F: Struct. Biol. Commun.* **73**, 174-183 (2017).

8. Jacobsen, C. *X-ray Microscopy.* (Cambridge Univ. Press, 2019).

9. Paunesku, T. et al. X-ray fluorescence microprobe imaging in biology and medicine. *J. Cel. Biochem.* **99**, 1489-1502 (2006).

10. Fahrni, C. J. Biological applications of X-ray fluorescence microscopy: exploring the subcellular topography and speciation of transition metals. *Curr. Opin. Chem. Biol.* **11**, 121-127 (2007).

11. Matsuyama, S. et al. Elemental mapping of frozen-hydrated cells with cryo-scanning x-ray fluorescence microscopy. *X-Ray Spectrom.* **39**, 260–266 (2010).

12. Takman, P. A. C. et al. High-resolution compact X-ray microscopy. *J. Microsc.* **226**, 175–181 (2007).

13. Wachulak, P. et al. A compact 'water window' microscope with 60 nm spatial resolution for applications in biology and nanotechnology. *Microsc. Microanal.* **21**, 1214–1223 (2015).

14. Schneider, G. Cryo X-ray microscopy with high spatial resolution in amplitude and phase contrast. *Ultramicroscopy* **75**, 85–104 (1998).

15. Yamamoto, Y. & Shinohara, K. Application of X-ray microscopy in analysis of living hydrated cells. *Anat Rec.* **269**, 217–223 (2002).

16. Berglund, M. et al. Compact water-window transmission X-ray microscopy. *J. Microsc.* **197**, 268–273 (2000).

17. Bertilson, M. et al. Laboratory soft X-ray microscope for cryotomography of biological specimens. *Opt. lett.* **36**, 14 (2011).

18. Duke, E. et al. Biological applications of cryo-soft X-ray tomography. *J. Microsc.* **255**, 65–70 (2014).

19. Nugent, K. A. Coherent methods in the X-ray sciences. *Adv. Phys.* **59**, 1–99 (2010).

20. Rodenburg, J. & Maiden, A. Ptychography in *Handbook of Microscopy*. 819-904 (Springer, 2019).

21. Giewekemeyer, K. et al. Quantitative biological imaging by ptychographic X-ray diffraction microscopy. *Proc. Natl. Acad. Sci. U. S. A.* **107**, 529–534 (2010).

22. Shapiro, D. A. et al. Chemical composition mapping with nanometre resolution by soft X-ray microscopy. *Nat. Photonics* **8**, 765-769 (2014).

23. Jones, M. et al. Mapping biological composition through quantitative phase and absorption X-ray ptychography. *Sci. Rep.* **4**, 1-4 (2014).

24. Piazza, V. et al. Revealing the structure of stereociliary actin by X-ray nanoimaging. *ACS Nano* **8**, 12228–12237 (2014).

25. Rodriguez, J.A. et al. Three-dimensional coherent X-ray diffractive imaging of whole frozen-hydrated cells. *IUCrJ* **2**, 575-583 (2015).

26. Rose, M. et al. Quantitative ptychographic bio-imaging in the water window. *Opt. Express* **26**, 1237-1254 (2018).

27. McPherson, A. et al. Studies of multiphoton production of vacuum-ultraviolet radiation in the rare gases. *J. Opt. Soc. Am. B* **4**, 595-601 (1987).

28. Hädrich, S. et al. Single-pass high harmonic generation at high repetition rate and photon flux. *J. Phys. B* **49**, 17:172022 (2016).

29. Rothhardt, J. et al. Table-top nanoscale coherent imaging with XUV light. *J. of Opt.* **20**, 11:113001 (2018).

30. Seaberg, M. D. et al. Tabletop nanometer extreme ultraviolet imaging in an extended reflection mode using coherent Fresnel ptychography. *Optica* **1**, 39–44 (2014).

31. Tanksalvala, M. et al. Nondestructive, high-resolution, chemically specific 3D nanostructure characterization using phase-sensitive EUV imaging reflectometry. *Sci. Adv.* **7**, 5 (2021).

32. Loetgering, L. et al. Tailoring spatial entropy in extreme ultraviolet focused beams for multispectral ptychography. *Optica* **8**, 130-138 (2021).

33. Goldberger, D. et al. Spatiospectral characterization of ultrafast pulse-beams by multiplexed broadband ptychography. *Opt. Express* **29**, 32474-32490 (2021).

34. Tadesse, G. K. et al. Wavelength-scale ptychographic coherent diffractive imaging using a high-order harmonic source. *Sci. Rep.* **9**, 1-7 (2019).

35. Eschen, W. et al. Material-specific high-resolution table-top extreme ultraviolet microscopy. *Light: Sci. & Appl.* **11**, 1-10 (2022).

36. Gardner, D. et al. Subwavelength coherent imaging of periodic samples using a 13.5 nm tabletop high-harmonic light source. *Nat. Photonics* **11**, 259–263 (2017).

37. Attwood, D & Sakdinawat, A. *Soft X-Rays and Extreme Ultraviolet Radiation.* (Cambridge Univ. Press, 2nd Ed 2017).

38. Baksh, P. et al. Quantitative and correlative extreme ultraviolet coherent imaging of mouse hippocampal neurons at high resolution. *Sci. Adv.* **6**, eaaz3025 (2020).

39. Kirz, J. et al. Soft X-ray microscopes and their biological applications. *Q. Rev. Biophys*. **28**, 33-130 (1995).

40. Grebing, C. et al. Kilowatt-average-power compression of millijoule pulses in a gas-filled multi-pass cell, *Opt. Lett*. **45**, 6250-6253 (2020).

41. Klas, R. et al. Ultra-short-pulse high-average-power megahertz-repetition-rate coherent extreme-ultraviolet light source. *PhotoniX* **2**, 1-8 (2021).

42. Klas, R. et al. Generation of coherent broadband high photon flux continua in the XUV with a sub-two-cycle fiber laser. *Opt. Express* **28**, 6188-6196 (2020).

43. Guizar-Sicairos, M. et al. Role of the illumination spatial-frequency spectrum for ptychography. *Phys. Rev. B* **86**, 100103 (2012).

44. Ostrowski, S. G. et al. Preparing Biological Samples for Analysis by High Vacuum Techniques. *Microsc. Today* **17**, 48-53 (2009).

45. Howells, M. R. et al. An assessment of the resolution limitation due to radiation-damage in x-ray diffraction microscopy. *J. Electron Spectrosc. Relat. Phenom.* **170**, 1-3 (2009).

46. Nave, C. The achievable resolution for X-ray imaging of cells and other soft biological material. *IUCrJ* **7**.3 (2020).

47. Farmand, M. et al. Near-edge X-ray refraction fine structure microscopy. *Appl. Phys. Lett.* **110**, 6 (2017).

48. Steinberg, G. Hyphal growth: a tale of motors, lipids, and the Spitzenkorper. *Eukaryotic cell* **6**, 351-360 (2007).

49. Henke, B. L. et al. X-ray interactions: photoabsorption, scattering, transmission, and reflection at E= 50-30,000 eV, Z= 1-92. *At. Data Nucl. Data Tables* **54**, 181-342 (1993).

50. Clutterbuck, A. J. Aspergillus nidulans. in *Bacteria, Bacteriophages, and Fungi.* (eds. King, R. C.) 447-510 (Springer, Boston, MA., 1974).

51. Schultzhaus, Z. et al. Phospholipid flippases DnfA and DnfB exhibit differential dynamics within the A. nidulans Spitzenkörper. *Fungal Genet. and Biol.* **99**, 26-28 (2017).

52. Riquelme, M. & Sánchez-León, E. The Spitzenkörper: a choreographer of fungal growth and morphogenesis. *Curr. Opin. Microbiol.* **20**, 27-33 (2014).

53. Huang, X. et al. Optimization of overlap uniformness for ptychography. *Opt. Express* **22**, 12634–12644 (2014).

54. Loetgering, L. et al. Ptylab: a cross-platform inverse modeling toolbox for conventional and Fourier ptychography. in *Proceedings of the OSA imaging and Applied Optics Congress 2021* (Optical Society of America, 2021).

55. Loetgering, L. et al. zPIE: an autofocusing algorithm for ptychography. *Opt. lett.* **45**, 2030-2033(2020).

56. Harada, T. et al. High-exposure-durability, high-quantum-efficiency (> 90%) backside-illuminated soft-X-ray CMOS sensor. *Appl. Phys. Express* **13**, 016502 (2019).

57. Huang, X. et al. Fly-scan ptychography. *Sci. Rep.* **5**, 9074 (2015).

58. Maiden, A., Johnson, D. & Li, P. Further improvements to the ptychographicaliterative engine. *Optica* **4**, 736–745 (2017).

59. Thibault, P. & Menzel, A. Reconstructing state mixtures from diffraction measurements. *Nature* **494**, 68–71 (2013).

60. Enders, B. Development and Application of Decoherence Models in Ptycho-graphic Diffraction Imaging. (PhD thesis, Technical University Munic, 2016).

61. Yao, Y. et al. Broadband X-ray ptychography using multi-wavelength algorithm. *J. Synchrotron Radiat.* **28**, 309–317 (2021).

62. Odstrcil, M. et al. Ptychographic coherent diffractive imaging with orthogonal probe relaxation. *Opt. Express* **24**, 8360–8369 (2016).